\documentclass[aps,pre,reprint,groupedaddress,showpacs]{revtex4-1}

\usepackage{graphicx}
\usepackage{dcolumn}
\usepackage{bm}
\begin{document}


\title{Protein sliding and hopping kinetics on DNA}


\author{Michael C. DeSantis$^1$, Je-Luen Li$^2$, and Y. M. Wang$^{\ast1}$}
\affiliation{$^1$Department of Physics, Washington University in St. Louis, Saint Louis, Missouri 63130, USA\\$^2$D. E. Shaw Research, New York, New York 10036, USA}
%
%



\date{\today}

\begin{abstract}
Using Monte Carlo simulations, we deconvolved the sliding and hopping kinetics of GFP-LacI proteins on elongated DNA from their experimentally observed seconds-long diffusion trajectories.  Our simulations suggest the following results: (1) in each diffusion trajectory, a protein makes on average hundreds of alternating slides and hops with a mean sliding time of several tens of ms; (2) sliding dominates the root mean square displacement of fast diffusion trajectories, whereas hopping dominates slow ones; (3) flow and variations in salt concentration have limited effects on hopping kinetics, while \textit{in vivo} DNA configuration is not expected to influence sliding kinetics; furthermore, (4) the rate of occurrence for hops longer than 200 nm agrees with experimental data for EcoRV proteins.
\end{abstract}


\maketitle


\section{Introduction \label{introduction}}
\begin{figure}[b]
\includegraphics{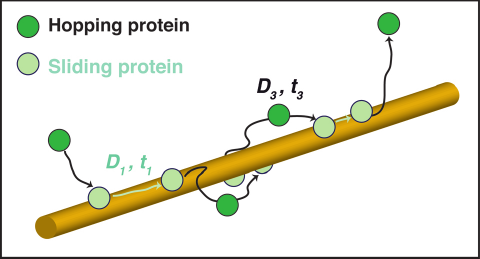}
\caption{\label{Fig1} (Color online) Schematics of a diffusion trajectory showing a protein initially binding to DNA, proceeding to slide (light disks) and hop (dark disks), and finally permanently dissociating from DNA.  This example diffusion trajectory has two discernible hops.}
\end{figure}

Timely target association of DNA-binding (DB) proteins is important for prompt cellular response to external stimuli using mechanisms such as gene regulation, DNA replication, and DNA repair.  The target association rates of DB proteins frequently deviate from the diffusion limit due to their interactions with nonspecific DNA via the process of facilitated diffusion \cite{Riggs1970,Marko2004,Wang2006}.  Facilitated diffusion mainly consists of two motions: sliding, where a protein diffuses along nonspecific DNA without losing contact, and hopping, where the protein jumps off DNA and undergoes 3D diffusion before reassociating to the same (Fig. \ref{Fig1}) or a different segment of DNA (referred to as intersegmental transfer). In this article, we regard events with long hopping distances, usually called jumping, as a form of hopping.  A DB protein may slide and hop many times on nonspecific DNA before reaching the target.  In order to quantify the effect of facilitated diffusion on DB proteins' target binding rate, how long a protein spends sliding on DNA (mean sliding time $\langle{t_1}\rangle$) and how fast it moves along DNA (sliding diffusion coefficient $D_1$) are two critical parameters for all calculations of \textit{in vitro} and \textit{in vivo} DNA geometries \  \cite{Marko2004,Wu2006PRL,Bruinsma2008,Mirny2008,Murugan2010,Spakowitz2010}.  
 
Single-molecule (SM) fluorescence imaging studies of DB proteins' Brownian diffusion along elongated DNA have obtained effective diffusion coefficients $D$ for the whole seconds-long diffusions (in this article we define each observed diffusion event between protein association and permanent dissociation to be a diffusion trajectory, and $t$ is the total time of the diffusion) \cite{Yanagida1999,Wang2006,Xie2006,Greene2006,Xie2007,Larson2007_2,Greene2007,Desbiolles2008,vanOijen2008,MirnyOijen2008,Stivers2008,Desbiolles2009,Scherer2009,Oijen2010}.  In the past, numerous studies had substituted $t$ and $D$ values in the place of $\langle{t_1}\rangle$ and $D_1$ in target binding rate and protein-nonspecific-DNA binding energy calculations since $\langle{t_1}\rangle$ and $D_1$ were not experimentally accessible  \cite{Wang2006,Xie2006,Xie2007,Bruinsma2008,Mirny2008,MirnyOijen2008,Elf2009,Murugan2010,Spakowitz2010}.  Since the extent of hopping involvement is unknown, it is dubious to use $t$ and $D$ values  for $\langle{t_1}\rangle$ and $D_1$.  Recent evidence suggests that these diffusion trajectories include both sliding and hopping: (1) the sliding time of DB proteins has been estimated to be milliseconds \cite{Revzin1990,Xie2007,Mirny2008,Elf2009};  (2) the sliding displacement has been estimated to be less than 50 bp \cite{Halford2005}, shorter than the displacements of whole diffusion trajectories of the reported DB proteins ($>$ 100 nm); (3) hops longer than 200 nm have been observed \cite{Desbiolles2008}. In order to obtain $\langle{t_1}\rangle$ and $D_1$ from experimental data, deconvolving sliding and hopping from individual diffusion trajectories is necessary.


\section{Simulations \label{simulations}}
Here we deconvolve sliding and hopping in a diffusion trajectory and obtain $\langle{t_1}\rangle$ and $D_1$ using (i) Monte Carlo simulations, (ii) experimental $D$ and $t$ values, and (iii) the following two relations (derived in \cite{Prove2010}):
\begin{eqnarray}
t   &=& N\langle{t_1}\rangle + N\langle{t_3}\rangle \mbox{ ,} \label{time} \\
2Dt &=& 2D_1N\langle{t_1}\rangle + 2D_3N\langle{t_3}\rangle \mbox{ ,} \label{2diffusions}
\end{eqnarray}
where $N$ is the mean number of sliding and hopping alternations in a diffusion trajectory, $D_3$ is the 3D diffusion coefficient of the protein, and $\langle{t_3}\rangle$ is the mean hopping time.  From hopping simulations we first determine $N$ and $\langle{t_3}\rangle$; then combining with experimental $D$ and $t$ values, $t_1$ and $D_1$ are obtained using Eqs. \ref{time} and \ref{2diffusions}.  

\begin{figure}[b]
\includegraphics{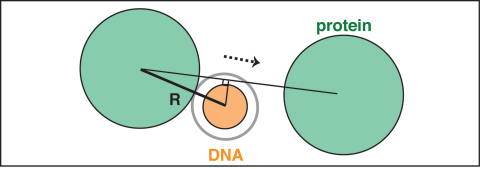}
\caption{\label{Fig2} (Color online) Determination of protein-DNA association.  The gray (open) circle marks the effective protein-DNA binding distance.  The protein moves ballistically between consecutive steps.}
\end{figure}

For each hopping simulation, a protein was initially positioned at the protein-center to DNA-center distance of $R$ = $r_{\rm DNA} + r_{\rm protein} + \Delta{r}$, where $r_{\rm DNA}$ = 1 nm is the DNA radius, $r_{\rm GFP-LacI}$ = 2.68 nm, and $\Delta{r} \approx$ 0.5 nm is an estimate of the protein-DNA binding distance (or location of the interaction potential minimum beyond which we consider no protein-DNA interactions) \cite{Joyeux2009_2,Victor2009}. The protein immediately dissociates from DNA and undergoes 3D diffusion until rebinding to DNA, at which time the position was recorded, or until the maximum number of steps of the hopping simulation was reached in which case the protein was assumed to have permanently dissociated and its diffusion trajectory was not used in subsequent data analysis. Figure \ref{Fig2} describes the criterion for determining whether a hopping protein collided with DNA.  For every step, the length of the perpendicular drawn from the center of the DNA to the line connecting the last two protein locations (dashed arrow) was calculated and if less than $R$, association occurred. The binding position was chosen to be the midpoint between the two protein locations.  We have modeled DNA as an infinite, rigid cylinder assuming 100\% probability for association upon protein-DNA  collision; the distance between the protein binding location and its origin denotes the hopping distance.

The simulation parameters were determined as follows.  The hopping simulation step size $\delta$, and step time $\tau$, are the collision distance and time, respectively \cite{Howard1993}.  At temperature $T$ = 294K, the instantaneous velocity of a protein of mass $m$ in solution is the root mean square (rms) velocity $\sqrt{\langle{v_x^2}\rangle}=\sqrt{k_BT/m}$ = $\delta/\tau$ = 6.02 m/s, where $k_B$ is the Boltzmann constant, $m$ = 67.5 kDa for a GFP-LacI monomer.  Using the Einstein-Stokes relation, $D_3={\delta}^2/(2\tau)$ = $k_BT/6\pi\eta{r} = 8.03\times10^7$ nm$^2$/s for GFP-LacI where the viscosity of water is $\eta = 10^{-3}$ N s/m$^2$ and the protein hydrodynamic radius $r$ is 2.68 nm assuming a typical protein density of 1.38 g/cm$^3$, we obtain $\delta = 2D_3/\sqrt{\langle{v_x^2}\rangle}$.  Therefore, $\delta = 0.267$ {\AA} and $\tau = 4.46$ ps.  Each simulation step in the $x$, $y$, $z$ dimensions was drawn from a Gaussian distribution with a mean of zero and a standard deviation of $\delta$.

The time limit for simulation of each GFP-LacI hop was $\approx$ 1 ms (or $2.1\times10^8$ steps), selected according to the following two estimations: (1) Since the observed diffusion of proteins on DNA is the combination of sliding and hopping with diffusion coefficients $D_1$ and $D_3$, respectively, the maximum total hopping time of a diffusion trajectory cannot exceed $N$$t_{3,max} = Dt/D_3$ when $D_{1} \approx 0$.  For GFP-LacI, $\langle{D}\rangle \approx 2\times10^4$ nm$^2$/s \cite{Wang2006} which dictates that $t_{3,max} \approx 0.25$ ms when $t$ is on the order of 1 s and using the low bound for $N$ of one hop per diffusion trajectory. Therefore, a hopping time limit of $t_{3,max}$ $\approx$ 1 ms for a single hop should be sufficiently long for all 3D diffusing proteins to return to DNA.  (2) A longer hopping time limit, such as 10 ms per hop (data not shown), results in additional proteins returning to DNA with individual hopping distances longer than $\sqrt{2\langle{D}\rangle{t}} = 200$ nm, a detectable distance in SM measurements that are usually used to separate single diffusion trajectories into segments free of large displacements for accurate $D$ analysis \cite{Wang2006,Desbiolles2008}.
 

\section{Results and Discussion \label{results}}
\begin{figure}[b]
\includegraphics{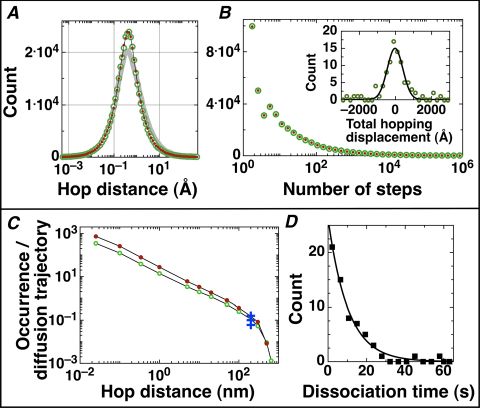}
\caption{\label{Fig3} (Color online) (A) Distributions of hopping distances along DNA for $\delta$ = 0.267 {\AA} and $R$ = 4.2 (green, open circles) and 10.2 nm (red dots), and hopping height for $R$ = 4.2 nm (gray line). (B) Distributions for number of steps per hop for $R$ = 4.2 and 10.2 nm. Inset, distribution for total hopping displacement per diffusion trajectory and Gaussian fit (solid line).  (C) Number of hops per diffusion trajectory longer than 0.25 \AA, and up to hops longer than 800 nm,  for $R$ = 4.2 and 10.2 nm.  The crosses are experimental data for EcoRV proteins, where the occurrence rate of hops per diffusion trajectory longer than 200 nm are 0.06, 0.1, and 0.16 (the 0.15 value was omitted for clarity) \cite{Desbiolles2008}. (D) GFP-LacI total diffusion time $t$ distribution (from experimental data in Ref. \cite{Wang2006}).  The mean of the exponential fit (solid line) is 10.4 s.}
\end{figure}

\begin{figure}[ht]
	\includegraphics{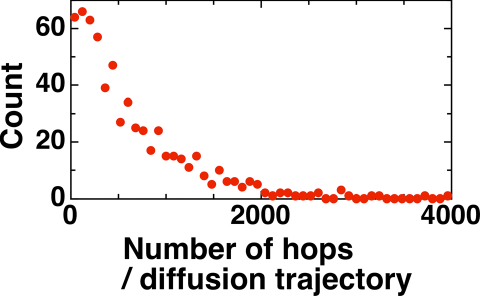}
	\caption{\label{Fig5} (Color online) Distribution of number of hops per diffusion trajectory. The results of $4\times10^5$ individual hopping simulations constitute a total of 763 protein diffusion trajectories such that 526 hops occur on average per trajectory.}
\end{figure}

For $4\times10^5$ GFP-LacI hopping simulations (maximum simulation time $t_{3,max} \approx$ 1 ms) with $\delta = 0.267$ {\AA} and $R = 4.2$ nm, 99.809\% of these trials resulted in the protein reassociating to DNA and thus the probability for a simulated hop to return to DNA is $P = 0.99809$.  The hopping characteristics are shown in Figs. \ref{Fig3}A and \ref{Fig3}B, in which the mean hopping distance along DNA is 3.37 {\AA} (median, 0.41 \AA), the mean hopping height (the maximum radial distance of the protein from DNA) is 4.93 {\AA} (median, 0.45 \AA), and the mean number of steps per hop is $4.97\times10^4$ (median, 5), yielding a mean hopping time $\langle{t_3}\rangle = 0.22$ $\mu$s.  The mean number of hops in a GFP-LacI diffusion trajectory is $N = 526$ obtained by dividing the total number of simulated hops of $4\times10^5$ by the total number of non-returned hopping events of 763; the distribution for the number of hops per diffusion trajectory is shown in Fig. \ref{Fig5}.  This set of values have been verified to converge with those from a larger simulation of $4\times10^6$ hops.  Specifically, $N$ values differ by 0.57\%. The inset of Fig. \ref{Fig3} shows the distribution of total hopping displacements in a diffusion trajectory with each data point simulated from 526 randomly selected hopping displacements.  The rms total hopping displacement per diffusion trajectory is 127.5 nm ($\sqrt{2D_3N\langle{t_3}\rangle}$), and the mean total hopping time is $N\langle{t_3}\rangle = 115$ $\mu$s.  Note that although shorter hopping distances, such as ones less than the base pair length of 0.34 {\AA}, do not carry direct biological significance nor do they noticeably disrupt sliding, they are important for correctly assessing rms total hopping displacement statistics in a diffusion trajectory.

We can also compute the `diffusion to capture' probability $P$ for a protein to return to DNA using a steady-state solution to the diffusion equation, incorporating a cutoff radial distance $c$ \cite{Howard1993}. Proteins released after the initial step at $b=4.22$ nm are either adsorbed at the DNA surface ($R=4.2$ nm) or escape beyond $c$ = $R+\sqrt{4D_3t_{3,max}}$. The probability is time-independent and given by
\begin{equation}
P = \frac{\log(c/b)}{\log(c/R)} = 0.99896 \mbox{ .}
\label{return}\vspace{-1mm}
\end{equation}
Imposing the same cutoff distance $c=551.2$ nm in subsequent simulations, we obtained $P=0.99865$, in near agreement with the analytical value above.

Having obtained $\langle{t_3}\rangle$ and $N$ from simulation, we now solve Eqs. \ref{time} and \ref{2diffusions} for $\langle{t_1}\rangle$ and $D_1$ from the experimentally measured values of $t$ and $D$.  With values of $D$ for GFP-LacI ranging from $2.3\times10^2$ to $1.3\times10^5$ nm$^2$/s \cite{Wang2006} and $t = 10.4$ s (Fig. \ref{Fig3}D), 
\begin{eqnarray}
\langle{t_1}\rangle &=& \frac{t}{N} - \langle{t_3}\rangle \approx \frac{t}{N} = 19.8 \mbox{ (ms)} \mbox{ ,}\\ 
D_1                 &\approx& D - \frac{D_3N\langle{t_3}\rangle}{t}\label{cnst} = D-\frac{9309}{10.4} \mbox{ (nm}^2}/{\mbox{s)} \mbox{ .}
\end{eqnarray}

The sliding time is several tens of ms and $D_1$ ranges from $\approx$ 0 for slow diffusion to $\approx D$ for fast diffusion.  The $\langle{D_1}\rangle$ for GFP-LacI is $9.1\times10^3$ nm$^2$/s using $\langle{D}\rangle$ of $10^4$ nm$^2$/s.  Since $D_1>0$, Eq. (\ref{cnst}) sets the lower bound of $D$ such that it must be greater than
$D_3N\langle{t_3}\rangle / t \approx 896 $ nm$^2$/s.
The rms total sliding displacement in a diffusion trajectory becomes longer than the rms total hopping displacement when $D>2ND_3t_3/t\approx 1790$ nm$^2$/s. 


Since our protein-nonspecific-DNA binding distance is an estimate, we have carried out simulations with $\Delta r$ ranging from $0.5$ to $6.5$ nm (corresponding to protein-DNA distances $R$ of 4.2 and 10.2 nm, respectively).  Comparing the $R$ = 10.2 nm results to the $R$ = 4.2 nm results, the distributions for hopping distances (Fig. \ref{Fig3}A) and hopping times (Fig. \ref{Fig3}B) are similar, although the mean hopping distance reduces to 2.82 \AA, the mean number of steps per hop reduces to $3.23\times10^4$, and the mean number of hops $N$, doubles to 1101. Solving for $\langle{t_1}\rangle$ and $D_1$ at $R=10.2$ nm, we found $\langle{t_3}\rangle$ = 0.14 $\mu$s, $N\langle{t_3}\rangle$ = 154 $\mu$s, $\langle{t_1}\rangle$ =  9.4 ms (approximately half of the value for $R$ = 4.2 nm), and $D_1$ to be similar to the previously calculated value for $R$ = 4.2 nm.  Given that the sliding and hopping values at $R$ = 4.2 and 10.2 nm are close, our method and results can be safely applied to most DB protein-DNA binding distances.  


To investigate hopping distances within a diffusion trajectory, Fig. \ref{Fig3}C shows the distribution of the number of hops per diffusion trajectory longer than a finite hopping distance, ranging from 0.25 {\AA} to 800 nm, for $R$ = 4.2 and 10.2 nm.  For the 4.2 nm results, 3.37 hops in a diffusion trajectory were longer than 5 nm, and 11\% of diffusion trajectories had a hop longer than 200 nm.  As expected, the results for 10.2 nm are approximately twice as large since $N$ is doubled.  The crosses represent EcoRV proteins, which have a comparable hydrodynamic radius of 2.66 nm (see Table \ref{Table1}), that were experimentally observed in different buffers to have hopped longer than 200 nm with reported occurrences ranging from 6 to 16\% per diffusion trajectory \cite{Desbiolles2008}. These observations are in agreement with our simulations results.  Furthermore, for hops longer than 300 nm and 500 nm, our observations agree with the reported values in Fig. 4A of Ref. \cite{Desbiolles2008}.  

\begin{table}[b]
\caption{\label{Table1}%
DB protein diffusion properties on elongated DNA.}
\begin{ruledtabular}
\begin{tabular}{lccr}
\textrm{Protein}&
\textrm{$r_{\rm protein}$ (nm)}&
\textrm{$\delta$ (\AA)}&
\textrm{$D$ (nm$^2$/s)}\\
\colrule
YFP-LacI, 2\footnote{2 indicates a dimer, and 4 indicates a tetramer.} & $3.13$ & $0.284$ & 4.6$\times$10$^{4}$ \cite{Xie2007}\\
GFP-LacI& $2.68$ & $0.267$ & 2.3$\times$10$^{2} - 1.3$$\times$10$^{5}$ \cite{Wang2006}\\ 
{EcoRV, 2} & $2.66$ & $0.262$ & 0.9$^{} - 2.5$$\times10^{4}$ \cite{Desbiolles2008}\\
EcoRV\footnote{Unknown molecular size due to unspecified/uncertain protein components and/or labels.} & $ $ & $$ & 3.1$\times$10$^{3}$ \cite{Desbiolles2009}\\ 
RNAP, 4\footnotemark[2] & $ $ & $$ & $6.1$$\times$10$^{3} - 4.3$$\times$10$^{5}$ \cite{Larson2007_2}\\
RNAP\footnotemark[2] & $$ & $ $ & $1.3$$\times$10$^{5}$ \cite{Heumann1988}, $\sim$10$^{4}$ \cite{Yanagida1999}\\
hOgg1 & $2.36$ & $0.247$ & $5.78$$\times$$10^{5}$ \cite{Xie2006}\\ 
p53 & $2.34$ & $0.246$ & $3.01$$\times$$10^{5}$ \cite{MirnyOijen2008}\\
UL42 &$2.63$ & $0.261$ & $5.1$$\times$$10^{3} - 2.2$$\times$$10^{4}$ \cite{vanOijen2008}\\ 
T7 gp5, 2 & $2.86$ & $0.272$ & $8.0$$\times$$10^{5} - 1.86$$\times$$10^{6}$ \cite{Oijen2010}\\
T7 gp5, 2 & $3.00$ & $0.278$ & $4.0$$\times$$10^{5}$ \cite{Oijen2010}\\
C-Ada &$1.77$ & $0.214$ & $1.3$$\times$$10^{6}$ \cite{Scherer2009}\\
\end{tabular}
\end{ruledtabular}
\end{table}

\begin{figure}[b]
\includegraphics{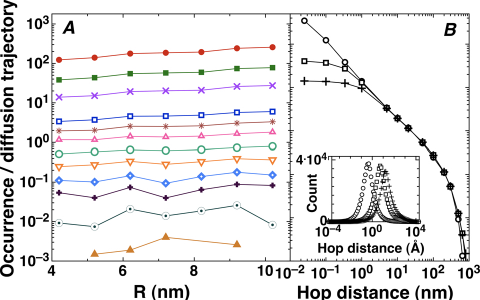}
\caption{\label{Fig4} (Color online) Distributions for number of hops per diffusion trajectory longer than 0.1, 0.34, 1, 5, 10, 20, 50, 100, 200, 300, 500, and 800 nm (top to bottom in A), (A) for $R$ ranging from 4.2 to 10.2 nm (left to right) and (B)  for $R$ = 4.2 nm and $\delta$ = 0.267 (circles), 3.4 (empty squares), and 10.2 {\AA} (crosses).  Inset, hopping distance distributions for the three $\delta$ values.}
\end{figure}

Other DB proteins may differ from GFP-LacI in their sizes, and thus $\delta$ and $R$.  Table \ref{Table1} lists DB proteins that can hop on DNA (instead of proteins that slide only \cite{Greene2006}) studied using SM fluorescence tracking methods on elongated DNA.  Despite the difference in $R$ by up to 1.26 nm, the $\delta$ values differ only by less than 0.07 \AA.  The effect of $R$ difference is considered in Fig. \ref{Fig4}A, in which the number of hops per diffusion trajectory longer than a finite distance, ranging from 0.1 {\AA} to 800 nm for $\delta$ = 0.267 {\AA} and $R$ from 4.2 to 10.2 nm are shown.  The number of hops per diffusion trajectory increases with $R$ moderately for all hopping distances, indicating that our hopping results are applicable to most observed DB proteins.  

The step size $\delta$ in the current approach, based on microscopic Brownian random walk models, can be made larger or smaller for vastly different particle sizes.  Figure \ref{Fig4}B shows distributions of hopping distances for three $\delta$ values: 0.267, 3.4, and 10 {\AA} (we used $R = 4.2$ nm and $t_{3,max} \approx 1$ ms).  The distribution curves collapse when protein hopping distances are larger than $\delta$, indicating that the tail distribution of protein hopping probability has the same asymptotic form at long distances, in agreement with the solution to the diffusion equation \cite{Desbiolles2009_2}.  However, the mean hopping distance (Fig. \ref{Fig4}B inset; values are 3.37, 36, and 95 \AA), the mean number of hops $N$, in a trajectory (526, 42, and 14), and $\langle{t_3}\rangle$ (0.22, 3.1, and 9.2 $\mu$s) all depend on $\delta$ sensitively, as short-length scale motions dominate protein-DNA reassociation (Fig. \ref{Fig3}A).  This regime can not be accessed in the macroscopic theory, i.e., by solving the diffusion equation directly.  

When the protein-nonspecific-DNA association probability $p$, is not 100\%, e.g., due to rotation of the DNA-binding domain during large hops, hopping statistics and the subsequent sliding statistics will change.  For a low binding probability of $p = 10\%$, although on average, ten consecutive hops would be needed for reassociation, the mean number of association attempts will still be $N$. However, the effective mean hopping time $\langle{t_3^{\prime}}\rangle$, and the mean hopping distance are expected to increase while the effective number of hops per diffusion trajectory $N^{\prime}$, decreases since $t$ is held constant.  The effective total hopping time $N^{\prime}\langle{t_3^{\prime}}\rangle$, and the rms total hopping distance per diffusion trajectory should therefore remain constant.  The binding probability is thus inversely related to the effective mean sliding time $\langle{t_1^{\prime}}\rangle$, according to Eq. \ref{2diffusions} which for $p = 10\%$ results in a 10-fold increase in $\langle{t_1^{\prime}}\rangle$.  

When salt concentration varies, $p$ and $R$ will change, as will $D_3$ within a few angstroms of the DNA surface.  However, since $t$ remains $\approx N^{\prime}\langle{t_1^{\prime}}\rangle$ because $N^{\prime}\langle{t_3^{\prime}}\rangle \ll N^{\prime}\langle{t_1^{\prime}}\rangle$, the observed changes in $t$ with salt concentration are likely due to changes in the total sliding time rather than the total hopping time.  Consequently, changes in $t$ as a result of varying salt concentration are not indicative of hopping and should not be used to determine its presence in diffusion trajectories, in disagreement with Refs. \cite{Xie2006,MirnyOijen2008,Mirny2008,vanOijen2008,Oijen2010}.  

Some studies use flow to elongate DNA and/or investigate hopping properties of DB proteins \cite{Xie2006,Larson2007,MirnyOijen2008,Scherer2009,Oijen2010}.  Here we describe the effect of flow on hopping distances using the maximum reported flow rate in SM studies of 100 $\mu$m/s.  For our mean hopping time of $\langle{t_3}\rangle = 0.22$ $\mu$s, a typical dissociated protein is carried by flow a length 0.22 {\AA} along DNA; this distance is negligible compared to its mean hopping distance of 3.37 {\AA} (the total displacement of the protein from flow alone within a diffusion trajectory consisting of 526 hops will be 11.6 nm which is substantially less than the total hopping displacement of 127.5 nm observed for GFP-LacI and similarly other proteins, as shown above). On the other hand, for a trajectory that includes a 1 $\mu$m-long hop, which occurs once every 1000 diffusion trajectories, the hopping time is 6.22 ms and a protein is flown 622 nm along DNA.  This distance would be sufficiently large for the protein to be considered dissociated.

Our results suggest that for diffusion trajectories without large hops of longer than of order a few hundred nanometers, a protein is unlikely to have been ``washed out'' while those that include large hops, the protein may be. However, according to Fig. \ref{Fig3}C, the probability for such an event to occur is approximately one percent of all diffusion trajectories.

Furthermore, sliding kinetics are not expected to be drastically affected by DNA configuration since a protein remains in contact with nonspecific DNA and should not be subject to DNA condensation and coiling either \textit{in vivo} or \textit{in vitro}, contrary to hopping kinetics. The reported values for $D_1$ and $t$ can therefore be applied under \textit{in vivo} situations for better estimation of target binding rates.

\section{Conclusion \label{conclusion}}
In summary, this study analyzes DB proteins' hopping on elongated DNA to address sliding kinetics. While we have made several assumptions regarding the nature of protein association and modeling DNA, our study suggests that the observed sliding kinetics is a robust feature. Although hopping kinetics will change according to \textit{in vivo} conditions, the lower bound on $D$ for a typical DB protein should help future experiments in identifying the presence of hopping in protein diffusion trajectories with greater certainty.


\begin{acknowledgments}
We are grateful to Anders Carlsson for helpful discussions.  M.C.D. wishes to thank the National Institutes of Health for a predoctoral fellowship awarded under 5T90 DA022871.
\end{acknowledgments}


\begin{thebibliography}{32}%
\makeatletter
\providecommand \@ifxundefined [1]{%
 \@ifx{#1\undefined}
}%
\providecommand \@ifnum [1]{%
 \ifnum #1\expandafter \@firstoftwo
 \else \expandafter \@secondoftwo
 \fi
}%
\providecommand \@ifx [1]{%
 \ifx #1\expandafter \@firstoftwo
 \else \expandafter \@secondoftwo
 \fi
}%
\providecommand \natexlab [1]{#1}%
\providecommand \enquote  [1]{``#1''}%
\providecommand \bibnamefont  [1]{#1}%
\providecommand \bibfnamefont [1]{#1}%
\providecommand \citenamefont [1]{#1}%
\providecommand \href@noop [0]{\@secondoftwo}%
\providecommand \href [0]{\begingroup \@sanitize@url \@href}%
\providecommand \@href[1]{\@@startlink{#1}\@@href}%
\providecommand \@@href[1]{\endgroup#1\@@endlink}%
\providecommand \@sanitize@url [0]{\catcode `\\12\catcode `\$12\catcode
  `\&12\catcode `\#12\catcode `\^12\catcode `\_12\catcode `\%12\relax}%
\providecommand \@@startlink[1]{}%
\providecommand \@@endlink[0]{}%
\providecommand \url  [0]{\begingroup\@sanitize@url \@url }%
\providecommand \@url [1]{\endgroup\@href {#1}{\urlprefix }}%
\providecommand \urlprefix  [0]{URL }%
\providecommand \Eprint [0]{\href }%
\@ifxundefined \urlstyle {%
  \providecommand \doi  [0]{\begingroup \@sanitize@url \@doi}%
  \providecommand \@doi [1]{\endgroup \@@startlink {\doibase
  #1}doi:\discretionary {}{}{}#1\@@endlink }%
}{%
  \providecommand \doi  [0]{doi:\discretionary{}{}{}\begingroup
  \urlstyle{rm}\Url }%
}%
\providecommand \doibase [0]{http://dx.doi.org/}%
\providecommand \Doi [0]{\begingroup \@sanitize@url \@Doi }%
\providecommand \@Doi  [1]{\endgroup\@@startlink{\doibase#1}\@@Doi}%
\providecommand \@@Doi [1]{#1\@@endlink}%
\providecommand \selectlanguage [0]{\@gobble}%
\providecommand \bibinfo  [0]{\@secondoftwo}%
\providecommand \bibfield  [0]{\@secondoftwo}%
\providecommand \translation [1]{[#1]}%
\providecommand \BibitemOpen [0]{}%
\providecommand \bibitemStop [0]{}%
\providecommand \bibitemNoStop [0]{.\EOS\space}%
\providecommand \EOS [0]{\spacefactor3000\relax}%
\providecommand \BibitemShut  [1]{\csname bibitem#1\endcsname}%
\bibitem [{\citenamefont {Riggs}\ \emph {et~al.}(1970)\citenamefont {Riggs},
  \citenamefont {Bougeois},\ and\ \citenamefont {Cohn}}]{Riggs1970}%
  \BibitemOpen
  \bibfield  {author} {\bibinfo {author} {\bibfnamefont {A.~D.}\ \bibnamefont
  {Riggs}}, \bibinfo {author} {\bibfnamefont {S.}~\bibnamefont {Bougeois}}, 
  and\ \bibinfo {author} {\bibfnamefont {M.}~\bibnamefont {Cohn}},\ }\href@noop
  {} {\bibfield  {journal} {\bibinfo  {journal} {J. Mol.
  Biol.}\ }\textbf {\bibinfo {volume} {53}},\ \bibinfo {pages} {401}
  (\bibinfo {year} {1970})}\BibitemShut {NoStop}%
\bibitem [{\citenamefont {Halford}\ and\ \citenamefont
  {Marko}(2004)}]{Marko2004}%
  \BibitemOpen
  \bibfield  {author} {\bibinfo {author} {\bibfnamefont {S.~E.}\ \bibnamefont
  {Halford}} and\ \bibinfo {author} {\bibfnamefont {J.~F.}\ \bibnamefont
  {Marko}},\ }\href@noop {} {\bibfield  {journal} {\bibinfo  {journal} {Nucl.
  Acids Res.}\ }\textbf {\bibinfo {volume} {32}},\ \bibinfo {pages} {3040}
  (\bibinfo {year} {2004})}\BibitemShut {NoStop}%
\bibitem [{\citenamefont {Wang}\ \emph {et~al.}(2006)\citenamefont {Wang},
  \citenamefont {Austin},\ and\ \citenamefont {Cox}}]{Wang2006}%
  \BibitemOpen
  \bibfield  {author} {\bibinfo {author} {\bibfnamefont {Y.~M.}\ \bibnamefont
  {Wang}}, \bibinfo {author} {\bibfnamefont {R.~H.}\ \bibnamefont {Austin}}, 
  and\ \bibinfo {author} {\bibfnamefont {E.~C.}\ \bibnamefont {Cox}},\
  }\href@noop {} {\bibfield  {journal} {\bibinfo  {journal} {Phys. Rev.
  Lett.}\ }\textbf {\bibinfo {volume} {97}},\ \bibinfo {pages} {048302}
  (\bibinfo {year} {2006})}\BibitemShut {NoStop}%
\bibitem [{\citenamefont {Klenin}\ \emph {et~al.}(2006)\citenamefont {Klenin},
  \citenamefont {Merlitz}, \citenamefont {Langowski},\ and\ \citenamefont
  {Wu}}]{Wu2006PRL}%
  \BibitemOpen
  \bibfield  {author} {\bibinfo {author} {\bibfnamefont {K.~V.}\ \bibnamefont
  {Klenin}} \textit{et al.}, }\href@noop {}
  {\bibfield  {journal} {\bibinfo  {journal} {Phys. Rev. Lett.}\
  }\textbf {\bibinfo {volume} {96}},\ \bibinfo {pages} {018104} (\bibinfo
  {year} {2006})}\BibitemShut {NoStop}%
\bibitem [{\citenamefont {Hu}\ \emph {et~al.}(2008)\citenamefont {Hu},
  \citenamefont {Grosberg},\ and\ \citenamefont {Bruinsma}}]{Bruinsma2008}%
  \BibitemOpen
  \bibfield  {author} {\bibinfo {author} {\bibfnamefont {L.}~\bibnamefont
  {Hu}}, \bibinfo {author} {\bibfnamefont {A.~Y.}\ \bibnamefont {Grosberg}}, 
  and\ \bibinfo {author} {\bibfnamefont {R.}~\bibnamefont {Bruinsma}},\
  }\href@noop {} {\bibfield  {journal} {\bibinfo  {journal} {Biophys.
  J.}\ }\textbf {\bibinfo {volume} {95}},\ \bibinfo {pages} {1151}
  (\bibinfo {year} {2008})}\BibitemShut {NoStop}%
\bibitem [{\citenamefont {Wunderlich}\ and\ \citenamefont
  {Mirny}(2008)}]{Mirny2008}%
  \BibitemOpen
  \bibfield  {author} {\bibinfo {author} {\bibfnamefont {Z.}~\bibnamefont
  {Wunderlich}} and\ \bibinfo {author} {\bibfnamefont {L.~A.}\ \bibnamefont
  {Mirny}},\ }\href@noop {} {\bibfield  {journal} {\bibinfo  {journal} {Nucl.
  Acids Res.}\ }\textbf {\bibinfo {volume} {36}},\ \bibinfo {pages} {3570}
  (\bibinfo {year} {2008})}\BibitemShut {NoStop}%
\bibitem [{\citenamefont {Murugan}(2010)}]{Murugan2010}%
  \BibitemOpen
  \bibfield  {author} {\bibinfo {author} {\bibfnamefont {R.}~\bibnamefont
  {Murugan}},\ }\href@noop {} {\bibfield  {journal} {\bibinfo  {journal}
  {Biophys. J.}\ }\textbf {\bibinfo {volume} {99}},\ \bibinfo {pages}
  {353} (\bibinfo {year} {2010})}\BibitemShut {NoStop}%
\bibitem [{\citenamefont {de~la Rosa}\ \emph {et~al.}(2010)\citenamefont {de~la
  Rosa}, \citenamefont {Koslover}, \citenamefont {Mulligan},\ and\
  \citenamefont {Spakowitz}}]{Spakowitz2010}%
  \BibitemOpen
  \bibfield  {author} {\bibinfo {author} {\bibfnamefont {M.~A.~D.}\
  \bibnamefont {de~la Rosa}} \textit{et al.}, }\href@noop {} {\bibfield  {journal} {\bibinfo
  {journal} {Biophys. J.}\ }\textbf {\bibinfo {volume} {98}},\
  \bibinfo {pages} {2943} (\bibinfo {year} {2010})}\BibitemShut {NoStop}%
\bibitem [{\citenamefont {Harada}\ \emph {et~al.}(1999)\citenamefont {Harada},
  \citenamefont {Funatsu}, \citenamefont {Murakami}, \citenamefont {Nonoyama},
  \citenamefont {Ishihama},\ and\ \citenamefont {Yanagida}}]{Yanagida1999}%
  \BibitemOpen
  \bibfield  {author} {\bibinfo {author} {\bibfnamefont {Y.}~\bibnamefont
  {Harada}} \textit{et al.}, }\href@noop {} {\bibfield
  {journal} {\bibinfo  {journal} {Biophys. J.}\ }\textbf {\bibinfo
  {volume} {76}},\ \bibinfo {pages} {709} (\bibinfo {year} {1999})}\BibitemShut
  {NoStop}%
\bibitem [{\citenamefont {Blainey}\ \emph {et~al.}(2006)\citenamefont
  {Blainey}, \citenamefont {van Oijent}, \citenamefont {Banerjee},
  \citenamefont {Verdine},\ and\ \citenamefont {Xie}}]{Xie2006}%
  \BibitemOpen
  \bibfield  {author} {\bibinfo {author} {\bibfnamefont {P.~C.}\ \bibnamefont
  {Blainey}} \textit{et al.}, }\href@noop
  {} {\bibfield  {journal} {\bibinfo  {journal} {Proc. Natl.
  Acad. Sci. U.S.A.}\ }\textbf {\bibinfo
  {volume} {103}},\ \bibinfo {pages} {5752} (\bibinfo {year}
  {2006})}\BibitemShut {NoStop}%
\bibitem [{\citenamefont {Graneli}\ \emph {et~al.}(2006)\citenamefont
  {Graneli}, \citenamefont {Yeykal}, \citenamefont {Robertson},\ and\
  \citenamefont {Greene}}]{Greene2006}%
  \BibitemOpen
  \bibfield  {author} {\bibinfo {author} {\bibfnamefont {A.}~\bibnamefont
  {Gran\'{e}li}} \textit{et al.}, }\href@noop {}
  {\bibfield  {journal} {\bibinfo  {journal} {Proc. Natl.
  Acad. Sci. U.S.A.}\ }\textbf {\bibinfo
  {volume} {103}},\ \bibinfo {pages} {1221} (\bibinfo {year}
  {2006})}\BibitemShut {NoStop}%
\bibitem [{\citenamefont {Elf}\ \emph {et~al.}(2007)\citenamefont {Elf},
  \citenamefont {Li},\ and\ \citenamefont {Xie}}]{Xie2007}%
  \BibitemOpen
  \bibfield  {author} {\bibinfo {author} {\bibfnamefont {J.}~\bibnamefont
  {Elf}}, \bibinfo {author} {\bibfnamefont {G.-W.}\ \bibnamefont {Li}},  and\
  \bibinfo {author} {\bibfnamefont {X.~S.}\ \bibnamefont {Xie}},\ }\href@noop
  {} {\bibfield  {journal} {\bibinfo  {journal} {Science}\ }\textbf {\bibinfo
  {volume} {316}},\ \bibinfo {pages} {1191} (\bibinfo {year}
  {2007})}\BibitemShut {NoStop}%
\bibitem [{\citenamefont {Kim}\ and\ \citenamefont
  {Larson}(2007)}]{Larson2007_2}%
  \BibitemOpen
  \bibfield  {author} {\bibinfo {author} {\bibfnamefont {J.~H.}\ \bibnamefont
  {Kim}} and\ \bibinfo {author} {\bibfnamefont {R.~G.}\ \bibnamefont
  {Larson}},\ }\href@noop {} {\bibfield  {journal} {\bibinfo  {journal}
  {Nucl. Acids Res.}\ }\textbf {\bibinfo {volume} {35}},\ \bibinfo
  {pages} {3848} (\bibinfo {year} {2007})}\BibitemShut {NoStop}%
\bibitem [{\citenamefont {Gorman}\ \emph {et~al.}(2007)\citenamefont {Gorman},
  \citenamefont {Chowdhury}, \citenamefont {Surtees}, \citenamefont {Shimada},
  \citenamefont {Reichman}, \citenamefont {Alani},\ and\ \citenamefont
  {Greene}}]{Greene2007}%
  \BibitemOpen
  \bibfield  {author} {\bibinfo {author} {\bibfnamefont {J.}~\bibnamefont
  {Gorman}} \textit{et al.}, }\href@noop {} {\bibfield
  {journal} {\bibinfo  {journal} {Cell}\ }\textbf {\bibinfo {volume} {28}},\
  \bibinfo {pages} {359} (\bibinfo {year} {2007})}\BibitemShut {NoStop}%
\bibitem [{\citenamefont {Bonnet}\ \emph {et~al.}(2008)\citenamefont {Bonnet},
  \citenamefont {Biebricher}, \citenamefont {Port\'{e}}, \citenamefont
  {Loverdo}, \citenamefont {B\'{e}nichou}, \citenamefont {Voituriez},
  \citenamefont {Escud\'{e}}, \citenamefont {Wende}, \citenamefont {Pingoud},\
  and\ \citenamefont {Desbiolles}}]{Desbiolles2008}%
  \BibitemOpen
  \bibfield  {author} {\bibinfo {author} {\bibfnamefont {I.}~\bibnamefont
  {Bonnet}} \textit{et al.}, }\href@noop {} {\bibfield  {journal}
  {\bibinfo  {journal} {Nucl. Acids Res.}\ }\textbf {\bibinfo {volume}
  {36}},\ \bibinfo {pages} {4118} (\bibinfo {year} {2008})}\BibitemShut
  {NoStop}%
\bibitem [{\citenamefont {Komazin-Meredith}\ \emph {et~al.}(2008)\citenamefont
  {Komazin-Meredith}, \citenamefont {Mirchev}, \citenamefont {Golan},
  \citenamefont {van Oijen},\ and\ \citenamefont {Coen}}]{vanOijen2008}%
  \BibitemOpen
  \bibfield  {author} {\bibinfo {author} {\bibfnamefont {G.}~\bibnamefont
  {Komazin-Meredith}} \textit{et al.}, }\href@noop
  {} {\bibfield  {journal} {\bibinfo  {journal} {Proc. Natl. Acad. Sci. U.S.A.}\ }\textbf {\bibinfo
  {volume} {105}},\ \bibinfo {pages} {10721} (\bibinfo {year}
  {2008})}\BibitemShut {NoStop}%
\bibitem [{\citenamefont {Tafvizi}\ \emph {et~al.}(2008)\citenamefont
  {Tafvizi}, \citenamefont {Huang}, \citenamefont {Leith}, \citenamefont
  {Fersht}, \citenamefont {Mirny},\ and\ \citenamefont {van
  Oijen}}]{MirnyOijen2008}%
  \BibitemOpen
  \bibfield  {author} {\bibinfo {author} {\bibfnamefont {A.}~\bibnamefont
  {Tafvizi}} \textit{et al.}, }\href@noop {} {\bibfield
   {journal} {\bibinfo  {journal} {Biophys. J.}\ }\textbf {\bibinfo
  {volume} {95}},\ \bibinfo {pages} {L01} (\bibinfo {year} {2008})}\BibitemShut
  {NoStop}%
\bibitem [{\citenamefont {Porecha}\ and\ \citenamefont
  {Stivers}(2008)}]{Stivers2008}%
  \BibitemOpen
  \bibfield  {author} {\bibinfo {author} {\bibfnamefont {R.~H.}\ \bibnamefont
  {Porecha}} and\ \bibinfo {author} {\bibfnamefont {J.~T.}\ \bibnamefont
  {Stivers}},\ }\href@noop {} {\bibfield  {journal} {\bibinfo  {journal}
  {Proc. Natl. Acad. Sci. U.S.A.}\ }\textbf {\bibinfo {volume} {105}},\ \bibinfo {pages} {10791}
  (\bibinfo {year} {2008})}\BibitemShut {NoStop}%
\bibitem [{\citenamefont {Biebricher}\ \emph {et~al.}(2009)\citenamefont
  {Biebricher}, \citenamefont {Wende}, \citenamefont {Escud\'{e}},
  \citenamefont {Pingoud},\ and\ \citenamefont {Desbiolles}}]{Desbiolles2009}%
  \BibitemOpen
  \bibfield  {author} {\bibinfo {author} {\bibfnamefont {A.}~\bibnamefont
  {Biebricher}} \textit{et al.}, }\href@noop {}
  {\bibfield  {journal} {\bibinfo  {journal} {Biophys. J.}\ }\textbf {\bibinfo {volume} {96}},\ \bibinfo {pages} {L50}
  (\bibinfo {year} {2009})}\BibitemShut {NoStop}%
\bibitem [{\citenamefont {Lin}\ \emph {et~al.}(2009)\citenamefont {Lin},
  \citenamefont {Zhao}, \citenamefont {Jian}, \citenamefont {Farooqui},
  \citenamefont {Qu}, \citenamefont {He}, \citenamefont {Dinner},\ and\
  \citenamefont {Scherer}}]{Scherer2009}%
  \BibitemOpen
  \bibfield  {author} {\bibinfo {author} {\bibfnamefont {Y.}~\bibnamefont
  {Lin}} \textit{et al.}, }\href@noop {} {\bibfield  {journal} {\bibinfo
  {journal} {Biophys. J.}\ }\textbf {\bibinfo {volume} {96}},\
  \bibinfo {pages} {1911} (\bibinfo {year} {2009})}\BibitemShut {NoStop}%
\bibitem [{\citenamefont {Etson}\ \emph {et~al.}(2010)\citenamefont {Etson},
  \citenamefont {Hamdan}, \citenamefont {Richardson},\ and\ \citenamefont {van
  Oijen}}]{Oijen2010}%
  \BibitemOpen
  \bibfield  {author} {\bibinfo {author} {\bibfnamefont {C.~M.}\ \bibnamefont
  {Etson}} \textit{et al.}, }\href@noop {} {\bibfield  {journal} {\bibinfo  {journal} {Proc. Natl. Acad. Sci. U.S.A.}\ }\textbf
  {\bibinfo {volume} {107}},\ \bibinfo {pages} {1900} (\bibinfo {year}
  {2010})}\BibitemShut {NoStop}%
\bibitem [{\citenamefont {Li}\ \emph {et~al.}(2009)\citenamefont {Li},
  \citenamefont {Berg},\ and\ \citenamefont {Elf}}]{Elf2009}%
  \BibitemOpen
  \bibfield  {author} {\bibinfo {author} {\bibfnamefont {G.-W.}\ \bibnamefont
  {Li}}, \bibinfo {author} {\bibfnamefont {O.~G.}\ \bibnamefont {Berg}},  and\
  \bibinfo {author} {\bibfnamefont {J.}~\bibnamefont {Elf}},\ }\href@noop {}
  {\bibfield  {journal} {\bibinfo  {journal} {Nat. Phys.}\ }\textbf
  {\bibinfo {volume} {5}},\ \bibinfo {pages} {294} (\bibinfo {year}
  {2009})}\BibitemShut {NoStop}%
\bibitem [{\citenamefont {Revzin}(1990)}]{Revzin1990}%
  \BibitemOpen
  \bibfield  {author} {\bibinfo {author} {\bibfnamefont {A.}~\bibnamefont
  {Revzin}},\ }\href@noop {} {\emph {\bibinfo {title} {The biology of
  nonspecific \textsc{DNA} protein interactions}}}\ (\bibinfo  {publisher} {CRC
  Press},\ \bibinfo {address} {London},\ \bibinfo {year} {1990})\BibitemShut
  {NoStop}%
\bibitem [{\citenamefont {Gowers}\ \emph {et~al.}(2005)\citenamefont {Gowers},
  \citenamefont {Wilson},\ and\ \citenamefont {Halford}}]{Halford2005}%
  \BibitemOpen
  \bibfield  {author} {\bibinfo {author} {\bibfnamefont {D.~M.}\ \bibnamefont
  {Gowers}}, \bibinfo {author} {\bibfnamefont {G.~G.}\ \bibnamefont {Wilson}},
   and\ \bibinfo {author} {\bibfnamefont {S.~E.}\ \bibnamefont {Halford}},\
  }\href@noop {} {\bibfield  {journal} {\bibinfo  {journal} {Proc. Natl. Acad. Sci. U.S.A.}\ }\textbf
  {\bibinfo {volume} {102}},\ \bibinfo {pages} {15883} (\bibinfo {year}
  {2005})}\BibitemShut {NoStop}%
\bibitem [{Pro()}]{Prove2010}%
  \BibitemOpen
  \href@noop {} {\bibinfo  {journal} {The DB protein's displacement on
  \textsc{DNA}, $x$, contains two alternating diffusion displacements: 1D sliding
  displacement $x_1$, and 3D hopping displacement $x_3$.
  $x=\sum_{i=1}^{N}{x}_{1i}+\sum_{j=1}^{N}{x}_{3j},
  \langle{x^2}\rangle=\sum_{i=1}^{N}\langle{x_{1i}}^2\rangle$+$\sum_{j=1}^{N}
  \langle {x_{3j}}^2\rangle$+$\sum_{i,j=1}^{N}2\langle
  x_{1i}{\cdot}x_{3j}\rangle=N\langle{x_1}^2\rangle+N\langle{x_3}^2\rangle=2D_%
1N \langle{t_1}\rangle+2D_3N\langle{t_3}\rangle$. This relation has also been
  verified by simulations.}}\BibitemShut {Stop}%
\bibitem [{\citenamefont {Florescu}\ and\ \citenamefont
  {Joyeux}(2009)}]{Joyeux2009_2}%
  \BibitemOpen
\bibfield  {journal} {  }\bibfield  {author} {\bibinfo {author} {\bibfnamefont
  {A.-M.}\ \bibnamefont {Florescu}} and\ \bibinfo {author} {\bibfnamefont
  {M.}~\bibnamefont {Joyeux}},\ }\href@noop {} {\bibfield  {journal} {\bibinfo
  {journal} {J. Chem. Phys.}\ }\textbf {\bibinfo {volume}
  {130}},\ \bibinfo {pages} {015103} (\bibinfo {year} {2009})}\BibitemShut
  {NoStop}%
\bibitem [{\citenamefont {Dahirel}\ \emph {et~al.}(2009)\citenamefont
  {Dahirel}, \citenamefont {Paillusson}, \citenamefont {Jardat}, \citenamefont
  {Barbi},\ and\ \citenamefont {Victor}}]{Victor2009}%
  \BibitemOpen
  \bibfield  {author} {\bibinfo {author} {\bibfnamefont {V.}~\bibnamefont
  {Dahirel}} \textit{et al.}, }\href@noop {} {\bibfield
  {journal} {\bibinfo  {journal} {Phys. Rev. Lett.}\ }\textbf {\bibinfo
  {volume} {102}},\ \bibinfo {pages} {228101} (\bibinfo {year}
  {2009})}\BibitemShut {NoStop}%
\bibitem [{\citenamefont {Berg}(1993)}]{Howard1993}%
  \BibitemOpen
  \bibfield  {author} {\bibinfo {author} {\bibfnamefont {H.~C.}\ \bibnamefont
  {Berg}},\ }\href@noop {} {\emph {\bibinfo {title} {Random walks in
  biology}}}\ (\bibinfo  {publisher} {Princeton University Press},\ \bibinfo
  {address} {New Jersey},\ \bibinfo {year} {1993})\BibitemShut {NoStop}%
\bibitem [{\citenamefont {Ricchetti}\ \emph {et~al.}(1988)\citenamefont
  {Ricchetti}, \citenamefont {Metzger},\ and\ \citenamefont
  {Heumann}}]{Heumann1988}%
  \BibitemOpen
  \bibfield  {author} {\bibinfo {author} {\bibfnamefont {M.}~\bibnamefont
  {Ricchetti}}, \bibinfo {author} {\bibfnamefont {W.}~\bibnamefont {Metzger}},
   and\ \bibinfo {author} {\bibfnamefont {H.}~\bibnamefont {Heumann}},\
  }\href@noop {} {\bibfield  {journal} {\bibinfo  {journal} {Proc. Natl. Acad. Sci. U.S.A.}\ }\textbf
  {\bibinfo {volume} {85}},\ \bibinfo {pages} {4610} (\bibinfo {year}
  {1988})}\BibitemShut {NoStop}%
\bibitem [{\citenamefont {Loverdo}\ \emph {et~al.}(2009)\citenamefont
  {Loverdo}, \citenamefont {B$\acute{e}$nichou}, \citenamefont {Voituriez},
  \citenamefont {Biebricher}, \citenamefont {bonnet},\ and\ \citenamefont
  {Desbiolles}}]{Desbiolles2009_2}%
  \BibitemOpen
  \bibfield  {author} {\bibinfo {author} {\bibfnamefont {C.}~\bibnamefont
  {Loverdo}} \textit{et al.}, }\href@noop {} {\bibfield  {journal} {\bibinfo  {journal} {Phys. Rev.
  Lett.}\ }\textbf {\bibinfo {volume} {102}},\ \bibinfo {pages} {188101}
  (\bibinfo {year} {2009})}\BibitemShut {NoStop}%
\bibitem [{\citenamefont {Kim}\ \emph {et~al.}(2007)\citenamefont {Kim},
  \citenamefont {Dukkipati}, \citenamefont {Pang},\ and\ \citenamefont
  {Larson}}]{Larson2007}%
  \BibitemOpen
  \bibfield  {author} {\bibinfo {author} {\bibfnamefont {J.~H.}\ \bibnamefont
  {Kim}} \textit{et al.}, }\href@noop {} {\bibfield  {journal} {\bibinfo  {journal} {Nano. Res.
  Lett.}\ }\textbf {\bibinfo {volume} {2}},\ \bibinfo {pages} {185} (\bibinfo
  {year} {2007})}\BibitemShut {NoStop}%
\end{thebibliography}
\end{document}